\documentstyle[epsf,psfig,english,bibnorm]{lamuphys} 
\newcommand{\be}{\begin{eqnarray}}  
\newcommand{\ee}{\end{eqnarray}}  
  
\newcommand{\ww}{W(\nu,q^2)}

\newcommand{\spett}{P(k,E)}

\makeatletter  
\let\chapter\hid@chapter  
\makeatother  
\begin{document}  
  
\title{Nuclear $y$ and $x$ scaling\protect\footnote{Presented by G. West at the Elba Workshop on Electron-Nucleus Scattering, EIPC June 22-26 1998}}  
  
\author{Claudio\,Ciofi degli Atti\inst{1}, Dino\,Faralli\inst{1} and Geoffrey\ B. West\inst{2}}  
  
\institute{ Department of Physics, University of Perugia, and Istituto Nazionale di Fisica Nucleare,  
Sezione di Perugia, Via A. Pascoli, I--06100 Perugia, Italy  
\and  
Theoretical Division, T-8, MS B285, Los Alamos National Laboratory, Los Alamos, NM 87545, USA}  
  
\titlerunning{Nuclear $y$ and $x$ scaling}  
  
\maketitle  
  
\begin{abstract}  
The exact expression of the nuclear structure function describing inclusive lepton scattering off nuclei is recalled,  
and the basic approximations leading to non-relativistic and relativistic nuclear $y$ and $x$ scaling, are illustrated.  
The general and systematic features of y-scaling structure functions are pointed out, and a recently proposed novel  
approach to y-scaling, based on a global scaling variable, $y_G$, which incorporates the effect of the momentum 
 dependence of the nucleon removal energy, and therefore allows the  establishment of  a direct link between scaling  
functions and momentum distributions, is illustrated and applied to the analysis of a large body of data, pertaining  
to nuclei ranging from the deuteron to Nuclear Matter. A new type of scaling phenomenon, the nuclear $x$-scaling,   
based on a proper  analysis of  nuclear quasi-elastic data  in terms of the Bjorken scaling variable $x_B$, is shown  
to occur at high values of the four-momentum transfer $Q^2$; the usefulness of   nuclear  $x$ scaling is pointed out.   
\end{abstract}  
  
\section{The nuclear response in inclusive lepton-nucleus scattering}  
In inclusive lepton scattering off nuclei, $A(e,e')X$, the nuclear response, or  
structure function, $W(\nu ,q^2)$, which  represents the deviation of the cross section 
 from scattering from free nucleons, has the following exact form:  
\be  
W(\nu,q^2)=\int_{-\infty}^{+\infty}\frac{dt}{2\pi}e^{i(\nu-\frac  
{q^{2}}{2M})t}<\psi_{0}|\sum_{i,j=1}^{Z}  
\hat{Q}_i\hat{Q}_{j}e^{-i(H-E_0)t}  
 e^{-i(\frac{\vec{p}_i\vec{q}}{2M})t}|\psi_{0}>\label{1-1}  
\ee  
where $H$ is the  Hamiltonian of the target nucleus,  $\psi_0(E_0)$ its   ground state  
wave function (energy), ${\bf p}_i$ the momentum operator of nucleon $i$,  
 and  $q\equiv\vert\vec{q}\vert$ and $\nu$ $( Q^2 = {\bf q}^2 -{\nu}^2)$ the  
three-momentum and energy transfers ( for the sake of simplicity    
only  the Coulomb interaction has been considered).  
  
Due to the non commutativity of $H$ and $\frac{\vec{p}_i\vec{q}}{2M}$, Eq.(\ref{1-1}) cannot be  
evaluated exactly. As a matter of fact, one can write:  
\begin{eqnarray}  
H&=\sum_{i=1}^{A}\frac{{\bf p_{i}}^{2}}{2M}+ \sum_{i<j}v(ij)=&\nonumber\\  
 &=H_{A-1}+\sum_{1\neq j}v(1,j)+ \frac{{\bf p_{1}}^{2}}{2M} &   \label{1-2}  
\end{eqnarray}  
and it can be seen that in general:  
\be  
\left[\sum_{j\neq 1}v(1,j),\frac{\vec{p}_1\vec{q}}{2M}\right]\neq0.  
\ee  
Let us assume, for the time being, that $\left[\sum_{j\neq 1}v(1,j),\frac{\vec{p}_1\vec{q}}{2M}\right]=0$;  
if one then replaces the independent kinematical variables $q^2$ and $\nu$ by  $q^2$ and $x_0=\frac{q^{2}}{2M_A\nu}$, 
 one obtains from  Eq.(\ref{1-1}) \cite{west}:  
\be  
i(\nu-\frac{q^{2}}{2M})t=-\frac{i}{\nu}(x_0 -1)t  
\ee  
and:  
\be  
 \lim_{q^{2}\rightarrow\infty}\nu\ww\simeq  
 \sum_{i}Q_{i}^{2}\delta\left( x_{0}-1\right)  
 \ee  
which shows that at high momentum transfer the reduced function ${\nu}W(x_0,q^2)$ should scale to 
 a $\delta$ function. Such a phenomenon will be called {\em non relativistic nuclear $x$-scaling}.  
  
The non relativistic $y$-scaling is obtained by introducing the quantity \cite{west}:  
\be  
y_0\equiv\frac{2M\nu-q^2}{2Mq}  
\ee  
and, assuming again that $\left[H,\frac{\vec{p}_i\vec{q}}{2M}\right]=0$, one obtains:  
\be  
 \lim_{q^{2}\rightarrow\infty}q\ww=  
\int \frac{d\vec{k}_{\perp}}{2\pi}\int_{-\infty}^{\infty}dk_{\parallel}  
n(\vec{k}_{\perp},k_{\parallel})\cdot\delta(k_{\parallel}-y_0)\equiv f(y_0)  
\label{1-7}  
\ee  
where:  
\be  
\label{1-8}  
f(y_0)=\int n(\vec{k}_{\perp},k_{\parallel})d\vec{k}_{\perp}=2\pi  
\int_{|y_0|}^{\infty}n(k)k \,dk  
\ee  
is the {\em longitudinal momentum distribution} and n(k) is the conventional momentum  
distribution normalized such that:  
\be   
\int d^{3}k\,n(k)=\int_{-\infty}^{\infty}dy_0\,f(y_0)=1 \label{1-9}  
\ee  
  
The above picture has to be improved by considering that in reality   
$\left[H,\frac{\vec{p}_i\vec{q}}{2M}\right]\neq0$  
and that actual experimental data require the use of  relativistic kinematics.  
Both improvements will be implemented in the rest of the paper, by means of the   
relativistic plane wave impulse approximation, in which the "final state interaction"  
(FSI) term $\sum_{j\neq1}v(1,j)$ in Eq.(\ref{1-2}) is disregarded and the non relativistic quantity   
$(\vec{p}_i+\vec{q})/2M$ is replaced by its  relativistic analog   
$\sqrt{(\vec{p}_i+\vec{q})^2+M^2}-M$.  
One obtains:  
\be  
\ww=\sum_f\left|<\psi_{A-1}^{f},\vec{k}_N|\hat{O}|\psi_0>\right|^2\cdot  
\delta(\nu+M_A-E_N-E_{A-1}) \label{1-10}  
\ee  
where:  ${\bf k}_N = {\bf k}+\bf q$ and ${\bf k}\equiv{\bf k}_1$ are the momenta of the struck nucleon 
 after and before interaction, respectively,  
   $E_N=\sqrt{({\bf k}+{\bf q})^2+M^2}$ is the nucleon total energy, and $E_{A-1}=  
\sqrt{M_{A-1}^{*^2}+{\bf k}^2}$ the total energy of the final $A-1$ system. Starting  
from Eq.~(\ref{1-10}),  we will introduce and discuss the  {\it relativistic} nuclear $y$ and $x$ scaling.

\section{$y$-scaling}  
Eq. (\ref{1-10}), shows that the structure function $W({\nu},q^2)$ is governed by the nucleon  spectral function $\spett$  
which depends on the energy $(E)$ as well as the momentum of the nucleons.  
In the limit of $q^2\rightarrow\infty$ it can be shown that \cite{ciofi}:  
\be  
\lim_{q^{2}\rightarrow\infty} qW(\nu,q^2) = F(y)  
\ee  
\be  
F(y) = 2 \pi\int\displaylimits_{E_{min}}^{\infty}dE\int\displaylimits _{k_{min}(y,E)}^{\infty}k\,dk \spett= f(y)-B(y)\label{1-12}  
\ee  
where:  
\be  
B(y) = 2 \pi \int\displaylimits_{E_{min}}^{\infty}dE     
            \int\displaylimits _{\vert y\vert}^{k_{min}(y,E)}k\,dk  {P_{1} (k,E)}	  
\ee  
is the so called {\em binding correction}, with $P_{1}$being  that part of $P(k,E)$ generated by   
ground state correlations and $f(y)$ is given by Eq. \ref{1-8} with $y_0$ replaced by $y$. The 
 scaling variable $y$ is obtained (see below)from the relativistic energy conservation appearing in Eq. (\ref{1-10}) , and  
represents the longitudinal momentum of the nucleons having the minimal value of the removal energy \cite{ciofi}.  
The interesting quantity is   
$f(y)$, since its knowledge would provide $n(k)$ by inversion of Eq.(\ref{1-8}).  
Unfortunately, the extraction of $f(y)$ from the experimental data requires the knowledge of both the experimental  
asymptotic scaling  
function $F(y)$, and the theoretical binding correction $B(y)$.  
  
Over the past several years there have been vigorous theoretical and experimental efforts to explore $y$-scaling  
over a wide range of nuclei \cite{day}, using the relativistic scaling variable $y$, which, recently,  
has been shown to lead to scaling in the relativistic deuteron,within   
 both the light-front \cite{poly} and the Bethe-Salpeter \cite{kaptari} approaches.  
  
In Ref. \cite{ciofi} the asymptotic scaling function $F(y)$ has been obtained by an extrapolation procedure  
of existing data, so as  to get rid of FSI, and the longitudinal  momentum distribution $f(y)$ has been  
obtained by adding to $F(y)$ the binding correction $B(y)$ evaluated theoretically.  
In Ref. \cite{ciofiwest,faralli} it has been assumed that  
  $f(y)$  obtained 
in this way represents the experimental longitudinal momentum distributions, whose   general and  
universal features, that are essentially  
independent of the detailed dynamics, and are valid from A=2 through Nuclear  Matter,  have been pointed out:  
\begin{itemize}  
\item[i)] $f(0)$ decreases monotonically with $A$, from  $\sim 10 MeV^{-1}$  
when $A=2$ to $\sim 3 MeV^{-1}$ for heavy nuclei; moreover,  
for $y \sim 0$, $f(y) \sim (\alpha^{2} + y^{2})^{-1}$, with $\alpha$   
ranging from $\sim 45 MeV$ for $A=2$, to $\sim 140 MeV$ for $A=56$.  
\item[ii)] For $50MeV \le  \vert y\vert \le 200 MeV$,  $F(y) \sim e^{-a^{2}y^{2}}$ with  
$a$  ranging from $\sim 50 MeV$ for $A=2$, to $\sim 150 MeV$ for $A=56$.  
\item[iii)] For $\vert y\vert \ge 400 MeV$,  $f(y) \sim {C_2} e^{-b \vert y \vert}$, with $B$   
ranging from $2.5 \times 10^{-4} MeV ^{-1}$ for $A=2$,   
to $1 \times 10^{-3} MeV ^{-1}$ for $A=56$,   
and, most intriguingly, $b = 6 \times 10^{-3} MeV ^{-1}$,  \it{independent} of $A$.  
\end{itemize}  
  
The following simple  
form for $f(y)$ yields an excellent representation of these general  
features for all nuclei:  
 \begin{equation}  
f(y) = \frac{{C_1}e^{-a^{2}y^{2}}}{ \alpha^{2} + y^{2}}  + {C_2} e^{-b \vert y \vert} = f_0 + f_1  
\label{seven}  
\end{equation}  
The first term ($f_0$) dominates the small $y$-behavior, whereas the second term ($f_1$) dominates   
large $y$. The systematics of the first term are determined by the small and  
intermediate  momentum behaviour of the single particle wave function. For  
$\vert y\vert\le\alpha$ this can be straightforwardly  understood in terms of a zero range  
 approximation and is, therefore,  insensitive to details of the  
microscopic dynamics, or of a specific model. The small $k$ behavior of the  
single particle wave function is controlled by its separation energy,   
 $(Q\equiv M + M_{A-1} - M_A = E_{min})$ and is given by    
$(k^{2}+\alpha^{2})^{-1}$ so  $\alpha = (2\mu Q)^{1\over 2}$, $\mu$ being the  
 reduced mass of the nucleon.     
This agrees quantitatively very well with fits to the data summarized in (i) above.   
  
The most intriguing phenomenological characteristic of  
the data is that  {\it $f(y)$ falls off exponentially at large $y$ with a similar  
slope parameter for all  nuclei, including the deuteron}. Since (i) $b$ is almost the same for all nuclei   
{\it including} $A=2$, i.e.,  $f(y)$, at large $y$, appears to be simply the rescaled   
 scaling function of the deuteron; and (ii)    
$b (\approx 1.18 fm) \ll 1/\alpha_D (\approx 4.35 fm)$, it can be concluded that the term  
${C_2}e^{-b\vert y \vert}$ is related to   
the short range part of the deuteron wave function and reflects the universal nature   
of $NN$ correlations in nuclei.   
  
\begin{figure}[htbp] 
\centerline{\psfig{file=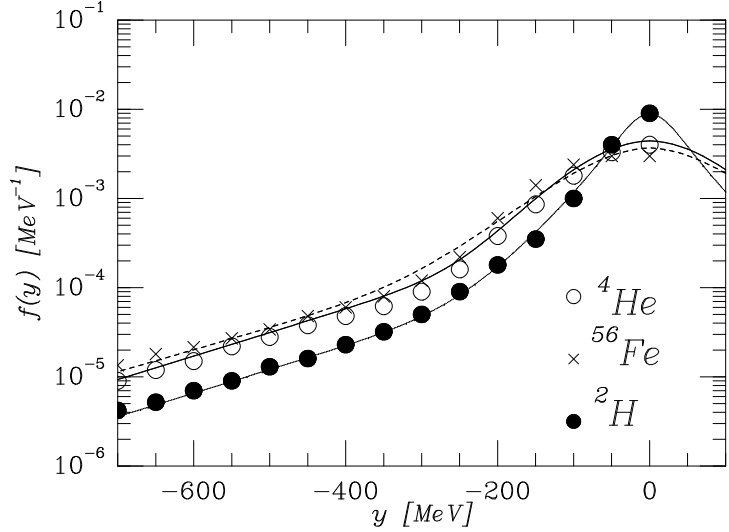,height=7cm,angle=270}} 
\caption[ ]{The longitudinal momentum distribution $f(y)$ for $^2H$ (dotted line), $^4He$ (full line) and $^{56}Fe$ (dashed line) corresponding to Eq. (\ref{seven}) with parameters given in Tab.1 . The points represent the "experimental" $f(y)$ obtained in Ref.\cite{ciofi}.} 
\label{fig1} 
\end{figure}

The remaining parameters, $C_1$ and $a$,  
can be related to $f(0)$ and the normalization condition,  
Eq.~(\ref{1-9}). Once this is done, there are no  
adjustable parameters for different nuclei.   
The intermediate range is clearly sensitive to $a$, the gaussian form   
being dictated by the shell model harmonic oscillator potential.  
Notice, however, that here the gaussian is modulated by the correct $\vert y\vert <\alpha$  
behaviour, namely $(y^2 + \alpha^2)^{-1}$, thereby ensuring the  
correct asymptotic wave function. The set of parameters of Eq.~(\ref{seven}) for various nuclei is presented in Tab.~1.  
As an example, Fig.~\ref{fig1} shows the  longitudinal scaling functions  $f(y)$  for $^2H$,  
 $^4He$ and $^{56}Fe$ extracted from the experimental data \cite{ciofi} compared to Eq.~(\ref{seven}).  
The fit is excellent.  
  
\begin{table}[htbp]  
\centering  
\caption{The parameters of Eq.~(\protect\ref{seven}) for various nuclei}  
\begin{tabular}{||l||c|c|c|c|c||}\hline \hline  
         &$C_1 [MeV]$&$\alpha [MeV]$&$a [MeV^{-1}]    $&$C_2 [MeV^{-1}]    $&$b [MeV^{-1}]  $\\ \hline\hline  
$^2H    $&$18     $&$45          $&$6.1\cdot 10^{-3}$&$2.5\cdot 10^{-4}$&$6\cdot 10^{-3}$\\ \hline  
$^3He   $&$41     $&$83          $&$7.1\cdot 10^{-3}$&$3.3\cdot 10^{-4}$&$6\cdot 10^{-3}$\\ \hline  
$^4He   $&$106    $&$167         $&$6.8\cdot 10^{-3}$&$6.5\cdot 10^{-4}$&$6\cdot 10^{-3}$\\ \hline  
$^{12}C $&$83     $&$166         $&$5.1\cdot 10^{-3}$&$5.7\cdot 10^{-4}$&$6\cdot 10^{-3}$\\ \hline  
$^{56}Fe$&$58     $&$138         $&$4.6\cdot 10^{-3}$&$6.2\cdot 10^{-4}$&$6\cdot 10^{-3}$\\ \hline\hline  
\end{tabular}  
\end{table}

With these observations it is now possible to understand the normalization and   
evolution of $f(y)$ with $A$. First note that Eq.~(\ref{1-8}) implies  
$f(0) = {1\over 2}\int {d^3 k} {n(k)\over k}  
=  \langle 1/2k \rangle$ and so  
is mainly sensitive to small momenta. Since  typical mean momenta vary from around 50 MeV for the deuteron up to   
almost 300 MeV for nuclear matter, it is clear  why $f(0)$ varies from around 10 for the deuteron to around   
2-3 for heavy nuclei. More specifically,  
since ${C_2}\ll {C_1}/\alpha^2$ and $f_1$  
falls off so  rapidly with $y$, the normalization integral,  
Eq.~(\ref{1-9}), is dominated by small $y$, i.e., by $f_0$. This leads to  
$f(0) \approx (\pi^{1/2} \alpha)^{-1} = (2\pi\mu Q)^{-1/2}$ which  
 gives an excellent fit to the $A$-dependence of $f(0)$.  
 Since $f(y)$ is  
constrained by the sum rule, Eq.~(\ref{1-9}), whose normalization is independent of   
the nucleus, a decrease in $f(0)$ as one changes the nucleus must be compensated   
for by a spreading of the curve for larger values of $y$.  {\it Thus, an understanding   
of $f(y)$ for small $y$ coupled with an approximately universal fall-off for large $y$,   
together with the constraint of the sum rule,  
leads to an almost model-independent understanding of the gross features of the data   
for all nuclei}.

To sum up, the "experimental" longitudinal momentum distribution can be thought of as  
the incoherent sum of a mean field shell-model contribution,  
$(f_{0})$, with the correct  
model-independent small $y$-behaviour built in, and a ``universal" deuteron-like  
correlation  contribution $(f_{1})$. Thus, the momentum distribution,  
$n(k)$, which is obtained from (\ref{1-8}), is also a sum of two contributions:  
$n=n_{0} + n_{1}$. This  
allows a comparison with results from many body calculations in which $n_{0}$ and   
$n_{1}$ have been separately calculated. Of particular relevance are not only the   
shapes of $n_{0}$ and $n_{1}$, but also their normalizations,   
$S_{0(1)}\equiv\int n_{0(1)}d^3k = \int f_{0(1)}dy$ which, theoretically,  
turn out to be, for $^4He$, $S_{0} \sim 0.8$ and   
$S_{1} \sim 0.2$~\cite{panda} whereas Eq.~(\ref{seven}) yields $S_0  
= 0.76$ and $S_1 = 0.24$.   
  
In order to minimize theoretical uncertainties arising in the subtraction of the 
 binding correction $B(y)$ a new scaling variable has been introduced in  
Ref. ~\cite{ciofiwest} wich in principle allows a determination of $f(y)$ free of theoretical contaminations.  
  
\begin{figure}[htbp] 
\centerline{\psfig{file=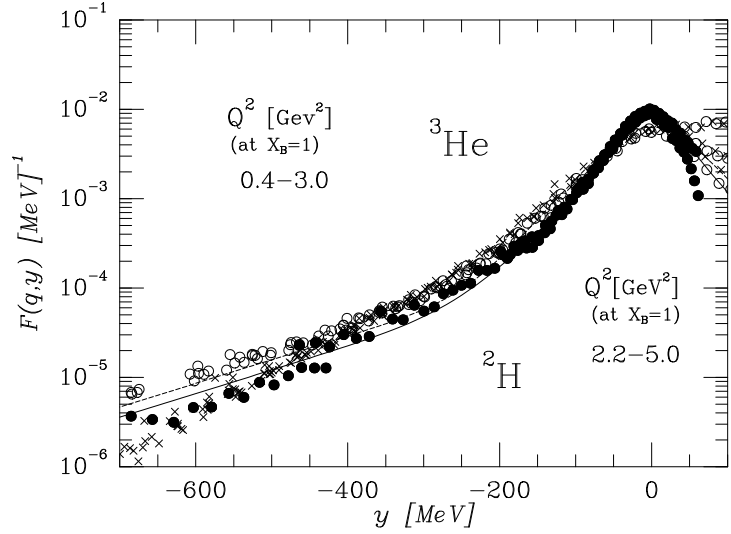,height=7cm,angle=270}} 
\caption[ ]{The experimental scaling function of $^3He$  plotted versus the usual, $y$, (crosses), and the global, $y_G$, (open dots) scaling variables,   compared with the scaling function of $^2H$ (full dots). The dashed and full lines are the calculated longitudinal momentum distributions of $^3He$ and $^2H$ respectively.} 
\label{fig2} 
\end{figure}

\begin{figure}[htbp] 
\centerline{\psfig{file=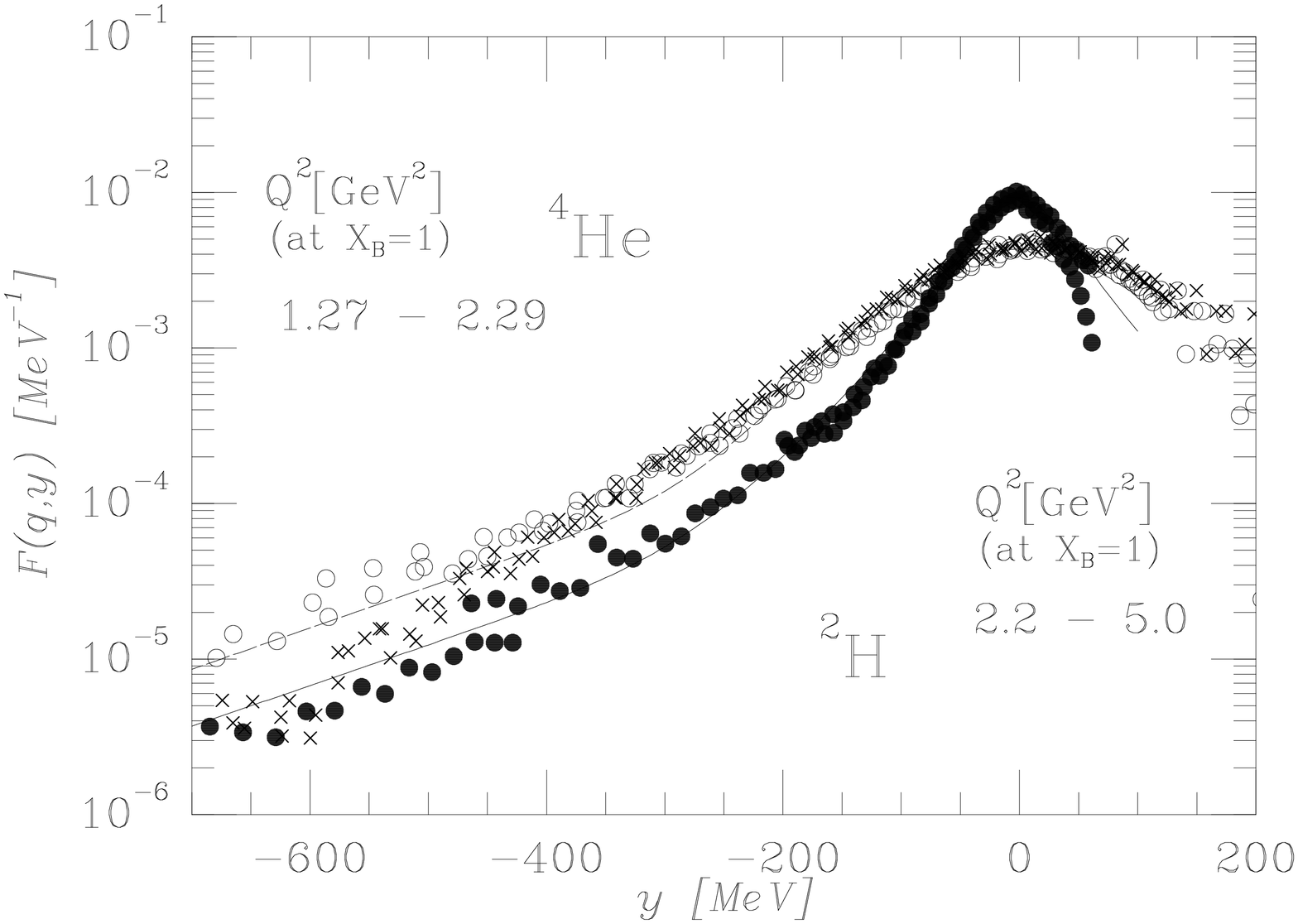,height=6.6cm,angle=270}} 
\caption[ ]{The same as in Fig.2 but for $^4He$.} 
\label{fig3} 
\end{figure}

\begin{figure}[htbp] 
\centerline{\psfig{file=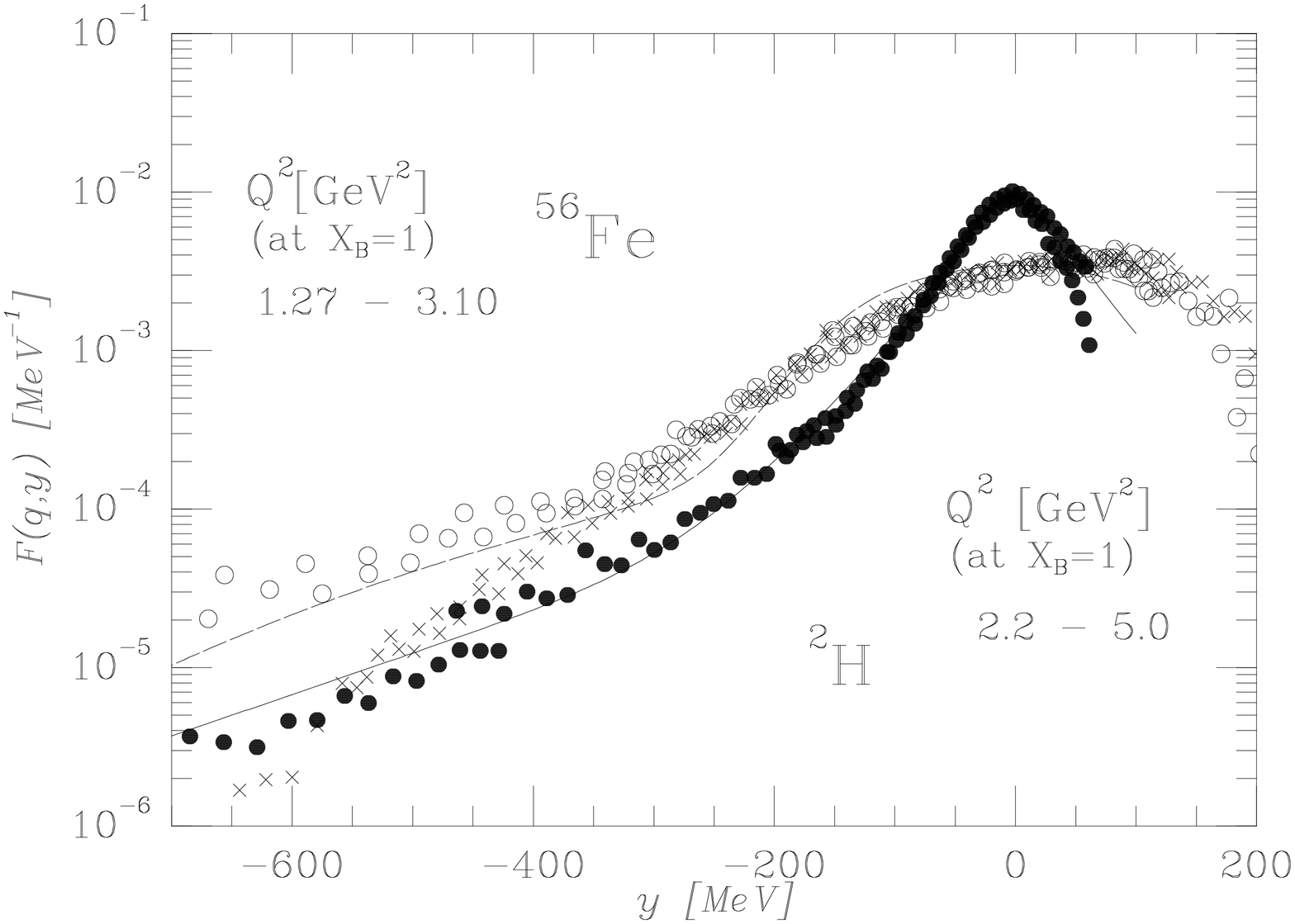,height=6.6cm,angle=270}} 
\caption[ ]{The same as in Fig.2 but for $^{56}Fe$.} 
\label{fig4} 
\end{figure}

\begin{figure}[htbp] 
\centerline{\psfig{file=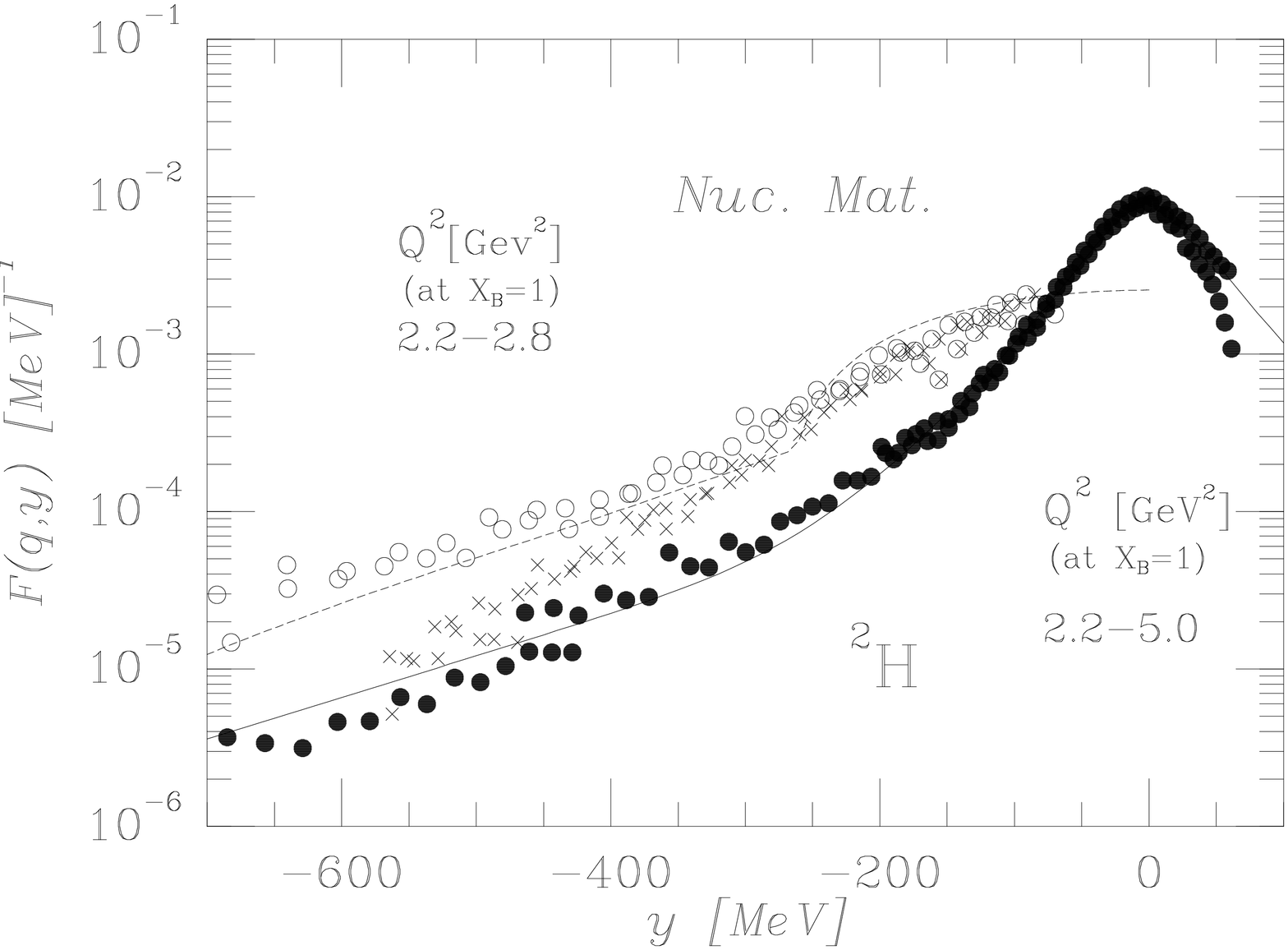,height=7cm,angle=270}} 
\caption[ ]{The same as in Fig.2 but for Nuclear Matter.} 
\label{fig5} 
\end{figure}

The usual scaling variable  $y$ is   
effectively obtained from energy conservation  
\be  
\nu + M_{A} = [(M_{A-1} + E_{A-1}^{*})^{2} + {\bf k}^2]^{1/2} + [M^{2} +({\bf k} + {\bf q})^{2}]^{1/2}  
\label{eleven}  
\ee  
by setting $k=|y|$, $\frac{{\bf k}\cdot {\bf q}}{kq} = 1$,   
and, most importantly, the excitation energy, $E_{A-1}^{*}=0$; thus, $y$ represents the nucleon   
longitudinal momentum of a nucleon having the {\it minimum value of the removal energy }  
$(E=E_{min},  E_{A-1}^{*} = 0)$. The minimum value of the nucleon momentum when  
 $q\to\infty$, becomes   
$k_{min}(y,E)=\vert y - (E - E_{min})\vert$. Only when $E=E_{min}$   
does $k_{min}(y, E) = \vert y \vert$, in which case $B = 0$ and $F(y)=f(y)$. However,   
the final spectator $(A-1)$ system can be  
left in all possible excited states,  including the continuum, so, in general,  
$E_{A-1}^{*} \ne 0$ and $E>E_{min}$, so $B(y) \ne 0$, and $F(y) \ne f(y)$. Thus,  
it is the dependence of $k_{min}$  on $E_{A-1}^{*}$ that gives rise to the binding  
effect, i.e. to the relation  $F(y) \ne f(y)$. This is an unavoidable defect of the usual approach  
to scaling; as a matter of fact, the longitudinal momentum is very different for weakly bound, shell 
 model nucleons (for which ${E_{A-1}^*} \sim 0-20 MeV$) and  strongly bound, correlated nucleons  
(for which  ${E_{A-1}^*} \sim 50-200 MeV$), so that at large values of $|y|$ the scaling function  
is not related to the longitunal momentum of those nucleons (the strongly bound, correlated ones) 
 whose contributions almost entirely exhaust the behaviour of the scaling function. In order to  
establish a general link between experimental data in different regions of the scaling variable,   
and longitudinal momentum components, one has to conceive a scaling variable which could equally  
well represent longitudinal momenta of both weakly bound and strongly bound nucleons.  
One can account for this in  
the following way.  The  
large $k$ and $E$ behaviours of  the Spectral Function  
are governed by configurations in which the high momentum of a  
correlated nucleon (1, say) is almost entirely balanced  by another   
nucleon (2, say), with the spectator $(A-2)$ system taking  only a small  
fraction of $k$, given by the $CM$ momentum of the pair $K_{CM}$ \cite{frank}. Within such a picture, one has   
 \cite{frank,simula}   
\begin{equation}  
E_{A-1}^{*} =  
\frac{A-2}{A-1}{\frac{1}{2M}}[{\bf {k}} -\frac {A-1}{A-2}{\bf {K}}_{CM}]^2  
\label{twelve}  
\end{equation}  
 which shows that the excitation energy of the residual nucleus depends both upon  
$\bf k$ and ${\bf K}_{CM}$; if the latter is set equal to zero, the average excitation energy  
for a given value of $k$ is $<{E_{A-1}^*}(k)>=\frac {A-2}{A-1}\frac {\vec k^{2}}{2M}$.  
 By replacing $E_{A-1}^*$ in  Eq.~(\ref{eleven}) with $\frac {A-2}{A-1}\frac {\vec k^{2}}{2M}$,  
the  deuteron-like   
scaling variable $y_{2}$ introduced in \cite{nucl-th} (see also \cite{ji},where the deuteron-like scaling  
variable was first introduced)  
  is obtained, representing the scaling variable pertaining to a ``deuteron" with mass  
 $\tilde M=2M-E_{th}^{(2)}$, where $E_{th}^{(2)} = \vert E_{A}\vert - \vert E_{A-2} \vert$.  
Such a scaling variable, however, has the unpleasant feature that the effect of the deuteron-  
like correlations are overestimated at low values of $y_2$ and , as a result, the correct,  
shell-model picture provided by the usual variable $y$ is lost.   
 When the  CM motion of the pair is taken into account, such a defect is cured.  
If the expectation value of Eq.~(\ref{eleven})  is evaluated with realistic spectral functions,  
for nuclei ranging from $^3He$ to Nuclear Matter, one obtaines (\cite{simula,faralli})  
\begin{equation}  
<E_{A-1}^{*}(k)> =  
\frac{A-2}{A-1}{\frac{1}{2M}}{\bf {k}}^2 +b_A-{c_A}\frac{{\bf k}^{2}}{2M}  
\label{tredici}  
\end{equation}  
with $b_A$ and $c_A$,  resulting   
 from the $CM$ motion of the pair,  
having values ranging  from $17 MeV$ to $43 MeV$ and $3.41 \times 10^{-1}$ to $1.66 \times 10^{-1}$,  
for $^3He$ and Nuclear Matter, respectively. Placing Eq.(\ref{tredici}) in Eq.(\ref{eleven}) and  
subtracting the value  of the average removal energy $<E>$ to counterbalance the value  (\ref{tredici})  
at low values of $y$, a new scaling variable is obtained which effectively takes into account  
the $k$-dependence of the excitation energy of the residual $A-1$ system, both at low and high  
values of $y$, unlike the usual scaling variable which completely disregards  $E_{A-1}^*$,  
and the scaling variable $y_2$, which overestimate the effects of deuteron-like correlations.  
In the kinematical region of existing experimental data this, $\it global$ scaling variable $y_G$  
reads as follows \cite{ciofiwest}  
 \begin{equation}  
y_{G}={\bigg\vert} -{{\tilde q}\over 2} +  
\left [{{\tilde q}^2\over{4}} - \frac{4  {\nu_{A}}^2 M^{2}- 
\rm{ W_A}^{4}}{\rm{ W_A}^{2}}\right ]^{1/2}{\bigg\vert}  
\label{quattordici}  
\end{equation}  
Here,  
$\nu_A = \nu + \tilde M$, $\tilde M = (2A-3)M/(A-1) - E_{th}^{(2)} -(b_A+2M^{2}c_A - <E>)$, ${\tilde q} = q -c_{A}\nu_{A}$,  
 and $\rm{W}_{A}^{2}   
= {\nu_A}^2 - {\vec q}^2 =\tilde M^{2} + 2 \nu  \tilde M - Q^{2}$.  
For the deuteron $E_{A-1}^{*}=0$, so $y_G\rightarrow y =  
 \vert -q/2 + [q^{2}/4 - (4   
{\nu_d}^{2} M^{2} - \rm{W_d}^{4})/{ \rm{W_d}^{2}}]^{1/2}\vert$    
with  $\nu_d =\nu + M_{d}$ and $\rm{W}_{d}^{2}   
= {\nu_d}^2 - {\vec q}^2 = M_{d}^{2} + 2 {\nu}  M_d - Q^{2}$.    
 For small values of $y_{G}$, such that ${(\frac{A-2}{A-1}{\frac{1}{2M}}{y}^2 +b_A-{c_A}\frac{{y}^{2}}{2M})}  
\ll <E>$,  the usual variable is recovered.  
 Thus $y_G$ interpolates between the correlation and the single particle regions.  
More importantly, however, since $k_{min}(q,\nu,E) \simeq \vert y_{G} \vert$,   
$B(y_G) \simeq 0$,  $F(y_G) \simeq f(y_G)$. Thus, plotting data in terms of  
$y_G$ allows a  {\it direct} determination of $f(y_G)$. One would therefore  
 expect from the above  analysis, the same behaviour of $f(y_G)$ at  
high  values of $y_G$ for both the deuteron and complex nuclei, unlike what happens with the usual  
scaling function $F(y)$, and the same shell-model behaviour at low values of $y$ as predicted by the 
 usual scaling variable. This is, indeed, the   
case, as exhibited in Fig.~(\ref{fig2}-\ref{fig5}), where the direct link between the scaling function  
$F(q,y_G)$ and the longitudinal momentum distributions is manifest.   
 
We can summarise our conclusions as follows:  
\begin{itemize}  
\item[i)] The general universal features of the $y$-scaling   
function have been identified and interpreted in terms of three contributions: a  
model-independent zero-range contribution, a ``universal" 2-nucleon correlation contribution and a  
mean field (shell-model) contribution;   
\item[ii)] The shape and evolution of the curve have been  
understood both  quantitatively and qualitatively on general grounds;    
\item[iii)] A global scaling variable which   
 incorporates the excitation  
energy of the $(A-1)$ system generated by correlations has been defined, which allows one to obtain the  longitudinal momentum  
distributions  directly  from the experimental data without introducing  
theoretical corrections.  
In terms of this variable the data strongly support the idea that the large $y$ behaviour  
in all nuclei is essentially nothing but a rescaled version of the  
deuteron.     
   
\end{itemize}

\section{$x$-scaling}  
  
It can be shown \cite{faralli} that the transition from the non relativistic $x_0$-scaling, discussed in section 1,  
to the relativistic one, can be achieved by using a minimal relativity Hamiltonian, i.e. $T_i=  
 \sqrt{\vec{p}_i^2+M^2}-M$, in which case the variable $x_0=q^2/2M\nu$ is replaced by the Bjorken scaling variable $x_0=Q^2/2M\nu$ .  
The central point, however,  is to understand how non relativistic $x_0$-scaling to a $\delta$ function is affected by the interaction effects due to $\left[H,\frac{\vec{p}_i\vec{q}}{2M}\right]\neq0$, and by the use of relativistic kinematics. This is discussed      
in  \cite{faralli} starting from Eqs.~(1) and (2). The results can be summarised as follows. Starting from the   relation:  
\be  
\nu W(x_B,q^2)=\frac{\nu}{q}qW(x_B,q^2)=\frac{\nu}{q}F(y,q^2)  
\ee  
and expressing $y$ through $x_{B}$,  the following equation, valid around $y\simeq 0$ $(x_{B}\simeq 1$) is found:  
\be  
\nu W(x_B,q^2)\simeq\frac{1}{(x_B-1)^2+\frac{\alpha^2}{M^2}}+ O(Q^{-2}) \label{nuw}  
\ee  
where we have used for $F(y,q^2)$ the expression $f(y)\propto\frac{1}{y^2+\alpha^2}$ (cf. Eq.~\ref{seven}).  
 
\begin{figure}[htbp] 
\centerline{\psfig{file=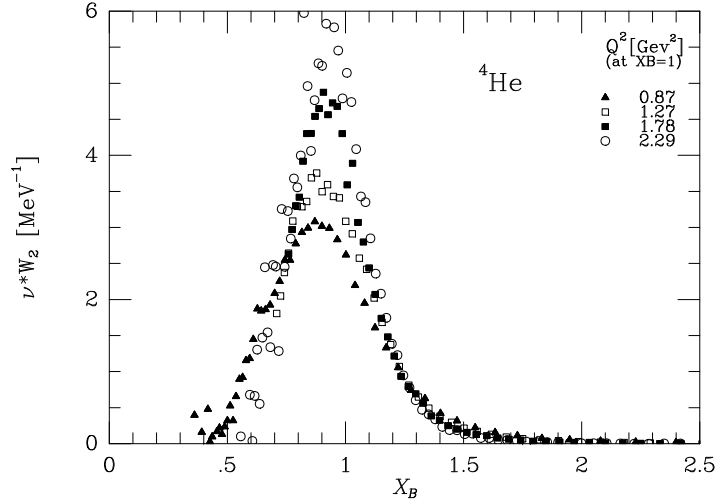,height=6.6cm,angle=270}} 
\caption[ ]{The reduced structure function $\nu\ww$ of $^4He$ plotted versus   the Bjorken variable $x_B$.} 
\label{fig6} 
\end{figure} 
 
\begin{figure}[htbp] 
\centerline{\psfig{file=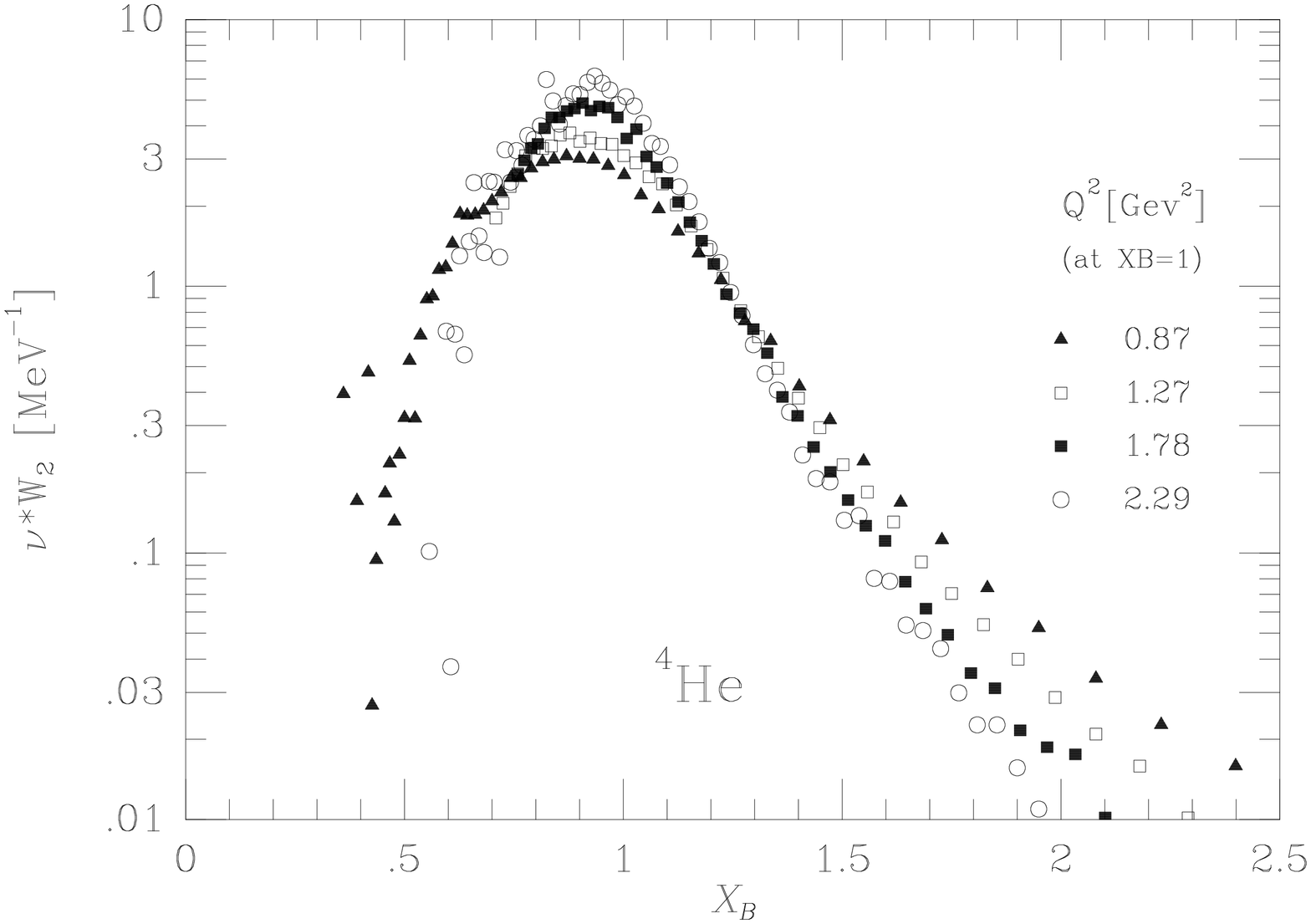,height=6.6cm,angle=270}} 
\caption[ ]{The log plot of Fig.6.} 
\label{fig7} 
\end{figure} 
 
\begin{figure}[htbp] 
\centerline{\psfig{file=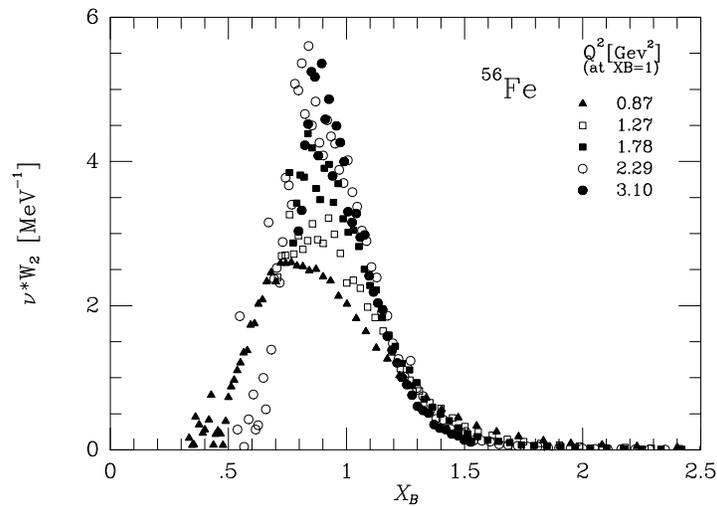,height=6.6cm,angle=270}} 
\caption[ ]{The same as in Fig.6 but for $^{56}Fe$.} 
\label{fig8} 
\end{figure} 
 
\begin{figure}[htbp] 
\centerline{\psfig{file=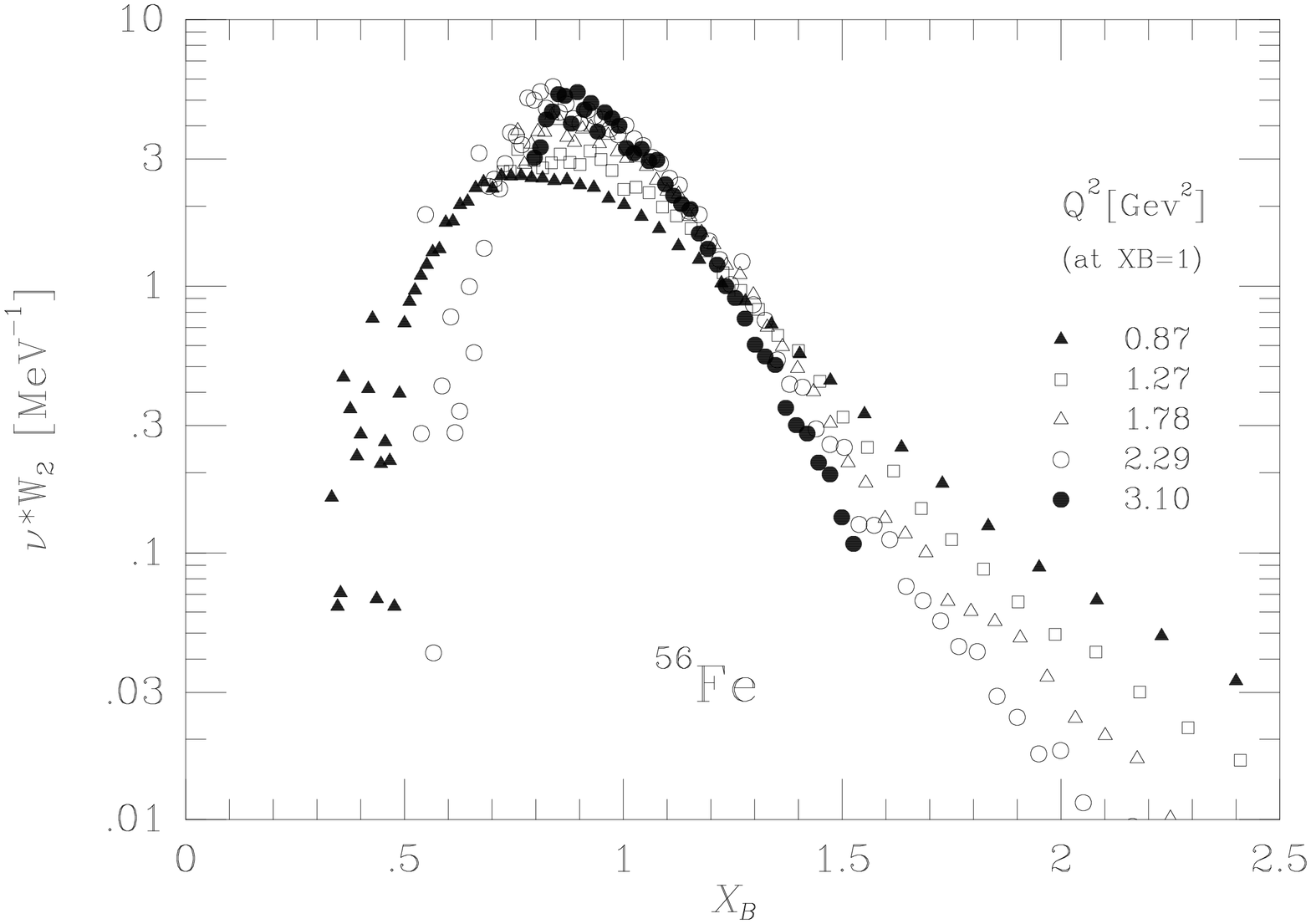,height=6.6cm,angle=270}} 
\caption[ ]{The log plot of Fig.8 .} 
\label{fig9} 
\end{figure}

It can be seen from Eq.~(\ref{nuw})  that by plotting the experimental data vs. $x_{B}$, one should not expect  
a $\delta$ function shape, as in  $x_0$-scaling, but a Lorentzian shape with a width decreasing  
with increasing $Q^2$  
and converging to a finite value ${\alpha^2}/{M^2}=2E_{min}/M$, when $Q^{2}\rightarrow\infty$.  
The plots   shown in Figs.~(\ref{fig6})-(\ref{fig9})  
seem  indeed to roughly exhibit   a Lorentzian behaviour, but   
experimental data at higher values of $Q^2$ would  be necessary to check the prediction of a saturating width.  
  
\section{Summary}  
  
The global $y$-scaling that we have discussed, by identifying the value of the global, $y_G$,  
scaling variable with the value of the nucleon longitudinal momentum, {\it  independently of  
the value of the nucleon removal energy}, allows one, unlike all previous approaches  
to $y$-scaling, to establish a direct relation between the scaling function and the  
longitudinal momentum distributions.  
 Nuclear $x$-scaling, i.e. scaling of inclusive {\it quasi elastic data on nuclei}, 
 when plotted vs. the Bjorken scaling variable $x_B$, has been shown to qualitatively occur;  
it can provide useful and complementary information on the nucleon momentum distributions.


\end{document}